\documentclass[%
onecolumn,
superscriptaddress,
nofootinbib,
amsmath,amssymb,
table
]{revtex4-2}

\usepackage{subfigure}
\usepackage{graphicx}
\usepackage{tikz}
\usepackage{booktabs} 
\usepackage{hyperref}

\usetikzlibrary{positioning}

\renewcommand{\thefootnote}{\fnsymbol{footnote}}

\definecolor{myblue}{RGB}{71,120,207}
\definecolor{mygreen}{RGB}{106,204,100}
\definecolor{myred}{RGB}{213,95,95}

\date{\today}

\makeatletter
\def\@email#1#2{%
 \endgroup
 \patchcmd{\titleblock@produce}
  {\frontmatter@RRAPformat}
  {\frontmatter@RRAPformat{\produce@RRAP{*#1\href{mailto:#2}{#2}}}\frontmatter@RRAPformat}
  {}{}
}%
\makeatother
\makeatletter
\renewcommand\paragraph{%
  \@startsection{paragraph}{4}{\z@}%
    {3.25ex \@plus1ex \@minus.2ex}%
    {-1em}%
    {\normalfont\normalsize\bfseries}%
}
\makeatother
\graphicspath{{figures/}}
\begin{document}

\preprint{AIP/123-QED}

\newcommand{\marker}[3]{
  \tikz[baseline=(X.base)]{
    \node [fill=#1!40,rounded corners] (X) {#2:};
  }
  {\color{#1!80!black}#3}
}

\newcommand{\tb}[1]{\marker{blue}{TB}{#1}}  

\title{Solvation Free Energies from Neural Thermodynamic Integration}%

\author{Bálint Máté\textsuperscript{*}}
\affiliation{Department of Computer Science, University of Geneva, Carouge, Switzerland}
\affiliation{Department of Physics, University of Geneva, Geneva, Switzerland}
\author{François Fleuret\textsuperscript{\dag}}%
\affiliation{Department of Computer Science, University of Geneva, Carouge, Switzerland}
\affiliation{Fundamental AI Research, Meta AI, Paris, France}
\author{Tristan Bereau\textsuperscript{\ddag}}%
\affiliation{Institute for Theoretical Physics, Heidelberg University, 69120 Heidelberg, Germany}
\affiliation{Interdisciplinary Center for Scientific Computing (IWR), Heidelberg University, 69120 Heidelberg, Germany}

\begin{abstract}
We present a method for computing free-energy differences using thermodynamic integration with a neural network potential that interpolates between two target Hamiltonians. The interpolation is defined at the sample distribution level, and the neural network potential is optimized to match the corresponding equilibrium potential at every intermediate time-step. Once the interpolating potentials and samples are well-aligned, the free-energy difference can be estimated using (neural) thermodynamic integration. To target molecular systems, we simultaneously couple Lennard-Jones and electrostatic interactions and  model the rigid-body rotation of molecules.  We report accurate results for several benchmark systems: a Lennard-Jones particle in a Lennard-Jones fluid, as well as the insertion of both water and methane solutes in a water solvent at atomistic resolution using a simple three-body neural-network potential.

\end{abstract}

\maketitle
\def\thefootnote{*}\footnotetext{balint.mate@unige.ch}
\def\thefootnote{\dag}\footnotetext{francois.fleuret@unige.ch}
\def\thefootnote{\ddag}\footnotetext{bereau@uni-heidelberg.de}

\section{Introduction}
Estimating free-energy differences is at the heart of understanding a wide range of physical, chemical, and biological processes, from protein folding and ligand binding to phase transitions in materials. These calculations offer invaluable insights into the stability of molecular conformations, the spontaneity of chemical reactions, and the mechanisms that drive phase changes \citep{beveridge1989free, gao2006mechanisms, mobley2017predicting, agarwal2021free}. The na\"ive way to compute free-energy differences would be to subtract two individual free energies. The issue with this approach is that the individual free energies are usually orders-of-magnitude larger than their difference. While theoretically sound, it requires extreme precision to predict a small quantity as the difference of two larger ones. Instead a variety of methods have been developed to compute free-energy differences directly. For instance, transfer free energies between different fluids  \cite{martin1997predicting, leroy2009interfacial}, solvation or hydration free energies \cite{mezei1987finite, straatsma1988free, helms1997free, martins2014prediction}, or protein-ligand binding \cite{brandsdal2003free, perozzo2004thermodynamics, deng2009computations, de2011free}. Estimating free energies or free-energy differences involves integrals over large configuration spaces and is usually evaluated using techniques from statistical physics such as thermodynamic integration (TI) and free-energy perturbation (FEP) \cite{mey2020best}. These computations  boil down to the task of sampling from equilibrium distributions and using the samples to perform numerical integration. To bypass the computation of such configurational integrals, it has been proposed to train machine-learning (ML) models on experimental or simulation labels to directly regress free energies \cite{riniker2017molecular, scheen2020hybrid, bennett2020predicting, rauer2020hydration, weinreich2021machine}. Another promising class of approaches seeks to replace classical sampling methods, such as Markov Chain Monte Carlo (MCMC) and molecular dynamics (MD), with techniques inspired by the machine learning community \citep{noe2019boltzmann, wirnsberger2020targeted, invernizzi2022skipping}. These methods leverage  generative models to produce i.i.d samples from their target distributions, enabling more effective computation of configurational integrals.

We build on NeuralTI \citep{mate2024neural}, an approach performing  TI using a neural-network potential, to estimate the free-energy difference between two arbitrary Hamiltonians. Thermodynamic integration estimates a free-energy difference between two Hamiltonians, $\mathcal H_0, \mathcal H_1$, defined on the same configuration space, via a one-parameter family of Hamiltonians $\mathcal H_\lambda$ interpolating between $\mathcal H_0$ and  $\mathcal H_1$. Given such an interpolation, the free-energy difference can then be written as an integral along the coupling interval $\Delta F_{0\to 1} = \int_0^1 \mathrm d \lambda  \langle  \partial \mathcal H_\lambda /\partial \lambda \rangle_\lambda$, where the brackets $\langle  .\rangle_\lambda$ denote the expectation value with respect to the Boltzmann distribution $e^{-\beta \mathcal H_\lambda}/{Z_\lambda}$. To numerically evaluate the TI integral, one usually discretizes the coupling variable  and generates  samples from the intermediate Boltzmann distributions at various values of $\lambda$. The idea behind NeuralTI  is to progressively transform samples between the endpoint distributions, learn the corresponding equilibrium $\mathcal H_\lambda$ with a neural network, and perform TI using the learnt $\mathcal H_\lambda$. This idea was first applied to a denoising diffusion model interpolating between the ideal gas and a Lennard-Jones (LJ) fluid in a periodic box \citep{mate2024neural}. This approach relied on the observation that the Hamiltonian of the ideal gas is trivial thus the corresponding Boltzmann distribution could be easily identified with the latent space of a diffusion model.

\begin{figure*}[!htb]
\begin{center}
\centerline{\includegraphics[width=\textwidth,trim={0cm 4cm 0cm 5cm},clip]{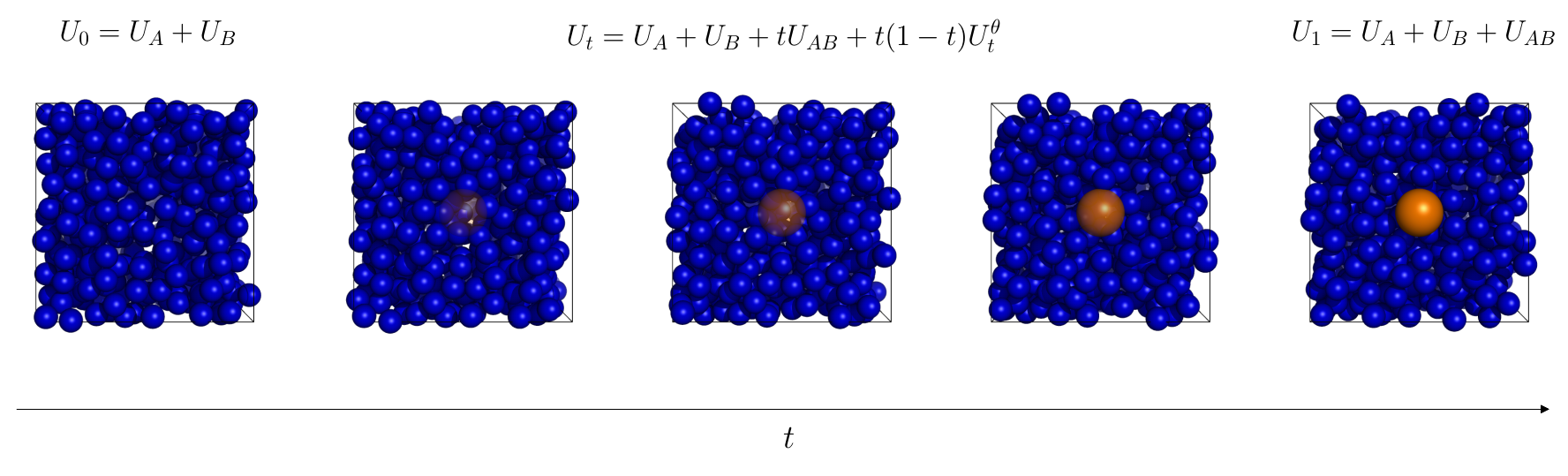}}
\vspace*{-3mm}
\caption{An interpolating family of distributions coupling a solute (brown) to the solvent (blue). The potentials $U_A,U_B$ and $U_{AB}$ denote the interactions within the solvent, within the solute and between the two components, respectively. The interpolation $U_t = U_A+U_B+tU_{AB}$ is often used in TI calculations to compute the free energy of the coupling of the solute, we include an additional trainable potential $t(1-t)U_t^\theta$ and train it to be the equilibrium potential at all intermediate time-slices.}
    \label{fig:summary}
\end{center}
\vskip -0.2in
\end{figure*}

In this work, we apply the same approach to a scenario where both endpoint Hamiltonians are not trivial, but represent  coupled and decoupled states of a solute--solvent system as illustrated in Fig. \ref{fig:summary}. This in particular means that we can no longer rely on the diffusion process for interpolation, but need to choose a different way of generating intermediate samples. To do this, we take inspiration from recent generalization of diffusion models \cite{lipman2022flow,albergo2023buildingnormalizingflowsstochastic} and construct samples $x_t\sim \rho_t$ as the geodesic interpolation between samples $x_0 \sim \rho_0, x_1 \sim \rho_1$.
Given these intermediate samples, the approach of neuralTI carries over to the current setting and we can optimize a neural network potential to match the equilibrium potential of $\rho_t$ at every $0<t<1$. In our experiments on chemical systems, we model both the Lennard-Jones and electrostatic interactions and couple them simultaneously as parametrized by the neural network.

The paper is structured as follows: Section \ref{sec:background} summarizes the relevant concepts that our work builds on. In Section \ref{sec:method} we argue that ideas from the machine-learning community generalizing diffusion models provide a natural framework for this setting \cite{lipman2022flow, albergo2023buildingnormalizingflowsstochastic, albergo2023stochasticinterpolantsunifyingframework} and describe our approach. Finally, in Section \ref{sec:experiments} we validate our methodology by computing solvation free energies on several systems: ($i$) an LJ solute in an LJ solvent, and the insertion of ($ii$) a water solute and ($iii$) a methane solute in a box of water, both at atomistic resolution. All free energies show excellent agreement with reference free energies.

\section{Background}
\label{sec:background}
We begin with fixing some notation. Depending on which arguments are important in the context, we will use $U_t(x)$ or $U(t,x)$ or simply $U$ to denote time-dependent potential functions and $\rho_t(x)$ for the corresponding Boltzmann probability density $\frac{1}{Z_t}e^{-\beta U_t(x)}$, with inverse temperature $\beta = 1/k_\mathrm{B}T$. The symbol $\nabla U$ is used for the spatial gradient. The isotropic Gaussian distribution, centered at the origin, with variance $\sigma^2$, is written as $\mathcal N_{\sigma^2}$ and the convolution of two functions, $f$ and $g$ as $f\star g$.
\paragraph*{Boltzmann densities and free energies}
Suppose a system, described by a Hamiltonian $\mathcal H$, is in thermal equilibrium with a heat reservoir. The likelihood of a particular microstate, described by coordinates $x$ and momenta $p$, is proportional to $e^{-\beta \mathcal H(x,p)}$. For the rest of this work we assume that all Hamiltonians are of the form $\mathcal H(x,p)=\sum_i \frac{p_i^2}{2m_i} + U(x)$, where $p_i$ and $m_i$ are the momenta and masses of the individual constituents of the system. Since the momenta only appear in the Hamiltonian through the first quadratic term, each component of each particle's momentum is normally distributed with variance $m/\beta$ and all the nontrivial behavior is described by the Boltzmann distribution of the positions $\rho(x) \propto e^{-\beta U(x)}$. The normalizing constant $Z=\int dx\,e^{-\beta U(x)}$ relates to the (Helmholtz) free energy, $F$, by $Z=e^{-\beta F}$. A usual quantity of interest is the free-energy difference between different Hamiltonians $F_{0\rightarrow 1} = F_1 -F_0 = \beta^{-1}(\log Z_0 -\log Z_1)$. Such free-energy differences indicate the relative stability of entire regions of conformational space---they are routinely used to predict solubility, the direction of reactions, or binding affinity.
\paragraph*{Thermodynamic Integration (TI)}
Thermodynamic integration \citep{kirkwood1935statistical} is a method for estimating the free-energy difference between two Hamiltonians relying on a family of potentials, $U_\lambda$, interpolating between $U_0$ and $U_1$.
{\allowdisplaybreaks
\begin{align}
    \beta \Delta F_{0\rightarrow 1}
    &=\log Z_0-\log Z_1 
    = -\int_0^1 \text{d}\lambda \, \partial_\lambda \log Z_\lambda  =  -\int_0^1 \text{d}\lambda \, \frac{1}{Z_\lambda}\partial_\lambda\left(\int \text{d}x \, \text{e}^{-\beta U_\lambda(x)}\right) \\
     &=  \beta\int_0^1 \text{d}\lambda \, \frac{1}{Z_\lambda}\left(\int \text{d}x \, \text{e}^{-\beta U_\lambda(x)} \partial_\lambda U_\lambda(x)\right) =  \beta \int_0^1 \text{d}\lambda \, \left\langle  \partial_\lambda U_\lambda \right\rangle_\lambda, \label{eq_TI} 
\end{align}}

where $\langle\partial_\lambda U_\lambda \rangle_\lambda$ denotes the expectation value of $\partial_\lambda U_\lambda$ under $\rho_\lambda = Z_\lambda^{-1}e^{-\beta U_\lambda}$. To numerically estimate the value of the last integral one needs an interpolating potential function $U_\lambda$ and samples from the corresponding interpolating equilibrium distributions $\rho_\lambda$. Usually one first fixes an interpolation $U_\lambda$ and performs Monte Carlo sampling or molecular dynamics (MD) simulations  at intermediate $\lambda$ values to obtain the estimate
\begin{equation}
    \Delta \hat F_{0\rightarrow 1} = \frac{1}{N}\sum_{\lambda_i} \, \mathbb E_{x\sim \rho_{\lambda _i}}\big[  \partial_\lambda U_\lambda(x)\big],
\end{equation}
where the sum is over $\lambda_i \in \{0,1/N,2/N,...,(N-1)/N\}$.

\paragraph*{Neural TI}
\citet{mate2024neural} proposed to parametrize a time-dependent energy function $U_t$ and use score matching (SM) to align the force $\nabla U_t$ with the scores of increasingly noised versions of the target Boltzmann distribution $e^{-\beta U_0}/Z_0$. As the noise level $\sigma(t)$ starts small at $t=0$, and is large at $t=1$, the trained $U_t$ provides an interpolating potential between the target and a tractable prior. Since samples from the corresponding Boltzmann distributions $e^{-\beta U_t} = \mathcal N_{\sigma_t^2} \star e^{-\beta U_0}$ are also (approximately) available, TI can be performed using $U_t$ to estimate the free-energy difference between a trivial and a non-trivial distribution. The prior and data densities were matched to the ideal gas and the Boltzmann distribution of the fully coupled LJ liquids, so as to associate the resulting free energy to the coupling of all the interactions. 
Traditional approaches to solvation free-energy calculations often consist of first fixing the interpolation in the space of potentials (e.g., linear coupling), to subsequently generate samples from Monte Carlo sampling or molecular dynamics simulations \cite{mey2020best}. Here instead, we first fix the sampling procedure and learn the interpolating family of potentials.

\paragraph*{Locally consistent potentials}
\label{sec:local_potentials}
The above strategy relies on learning a time-dependent potential $U_t$ corresponding to samples $x_t \sim \rho_t$. To perform TI, the samples $x_t$ and the energy function $\beta U_t$ need to be compatible with each other in the sense that $x_t \sim \frac{1}{Z_t}e^{-\beta U_t}$. An energy-based model trained with an SM objective can only fit an energy function up to an additive constant. Since we aim to learn a time-dependent family of energies, the degrees of freedom correspond to a time-dependent function defining the energy scale along the trajectory. Because  the boundary conditions are enforced by construction, the integral of this degree of freedom over the full time interval $[0,1]$ must vanish.

\begin{figure*}[h]
    \begin{center}
    \centerline{\includegraphics[width=\textwidth,trim={7cm 0cm 15cm 0cm},clip]{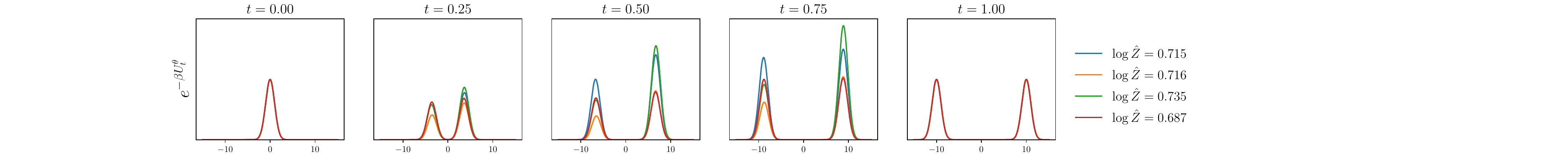}}
    \caption{Unnormalized interpolating densities $e^{-\beta 
U_t^\theta}$ learned from different initializations. The two endpoint potentials are given by a standard Gaussian at $t=0$ ($\log Z_0 =1$)  and the sum of two Gaussian densities at $t=1$ ($\log Z_1 =2$). The 4 different models learn 4 different relative weightings of the modes along the interpolation, but have a consistent prediction for the free-energy difference of $\log 2 \approx 0.69$.}
    \label{fig:toy_example}
    \end{center}
    \vskip -0.5in
\end{figure*}

A possible issue arises for a  domain that is disconnected or a target distribution that is multimodal: the additive constant becomes \emph{local}. Thus, strictly speaking, our model does not learn a function $U_t$ such that $e^{-\beta U_t}$ is globally proportional to $\rho_t$, but only locally proportional. In other words, it does not preserve the relative weights of the modes, but does preserve the relative likelihoods within the modes. \citet{song2019generative} solve this by learning a family of models with increasing noise levels, making use of the fact that for high perturbations the target distribution becomes connected. Alternatively, \citet{gutmann2010noise} overcome the issue by contrasting the data with a second distribution of known density. In the present work, the enforced boundary conditions alleviate the issue: All offsets introduced to the relative weighting of the modes also integrate to zero over the time interval $[0,1]$. See Figure \ref{fig:toy_example} for a one-dimensional illustration. 
While mathematically equivalent, from a numerical perspective, interpolations that yield lower variance of $\partial_t U_t$ are preferred for the TI-estimation of the free-energy difference. Whether the training converges to such a potential likely depends on the initialization scheme, the training dynamics and the explicit regularization of the variance of $\partial_t U_t$.

\section{Method}
\label{sec:method}
The goal of this work is to extend the neural TI framework to the estimation of free-energy differences between two non-trivial Hamiltonians. 
To this end, we consider a pair of potential functions, $U_0, U_1:\mathcal M \rightarrow \mathbb R$ on a manifold $\mathcal M$ representing the configuration space of some system. 

In this general setting, without further assumptions on the system, we can outline our proposed method.
\begin{enumerate}

    \item Choose a parametric family of interpolations between $U_0$ and $U_1$. We can always use $U(t,x) = (1-t)U_0(x) + tU_1(x)+t(1-t)U^\theta(t,x)$, where $U^\theta(t,x)$ is a parametric function, usually a neural network, with parameters $\theta$. The only requirement here is that the boundary conditions at $t\in \{0,1\}$ are satisfied.
    \item Define the sampling process of $\rho_t$---in other words, a way to construct samples $x_t \sim \rho_t$ given samples $(x_0,x_1) \sim (\rho_0,\rho_1)$. This highly depends on the geometry of $\mathcal M$. As long as $\mathcal M$ can be equipped with the structure of a Riemannian manifold, i.e., a metric $g \in C^\infty(\mathcal M,\text{Symm}^2(T^*\mathcal M))$, the geodesic interpolation with respect to $g$ is a straightforward choice.
    \item Given the two above-mentioned constructions, what remains is to optimize $U^\theta(t,x)$ until it approximates the equilibrium potential of $\rho_t$. In this work we do this by score matching.
\end{enumerate}

In our experiments, we focus on solvation free energies by modeling the intermolecular interactions between solute and solvent using common molecular-mechanics terms, except that we freeze intramolecular contributions by working with rigid molecules. All systems are embedded in a periodic box. For the particle system $\mathcal M$ will be the $3N$-torus, $\mathbb T^{3N}$ representing the positions of $N$ particles in a 3-torus. For the rigid molecular system, the configuration of each molecule is given by its position in $\mathbb T^{3}$ and its orientation, thus the configuration space of the full system is $\mathcal M = [\mathbb T^{3}\times SO(3)]^N$. 
\subsection*{Learning an interpolating potential}
\paragraph*{Sampling the interpolating densities}
Given two densities $\rho_0, \rho_1$, there are infinitely many possible interpolations between them. To obtain a unique and well-defined interpolation, one could impose additional conditions on the transport \cite{villani2009optimal,schrodinger1932theorie}. However, since this work focuses on computing the free-energy difference between the endpoints, all interpolations are equally valid, as the free energy is a state function. Consequently, our design choices regarding the interpolation are driven solely by considerations of computational simplicity. For what follows, we assume that we have access to samples $x_0\sim \rho_0  \propto e^{-\beta U_0}$ and $x_1 \sim \rho_1  \propto e^{-\beta U_1}$. We then define an interpolating family of distributions $\rho_t$ for $0<t<1$ by constructing a procedure to sample $x_t$ a linear interpolation between a pair of samples from the endpoint distributions \citep{albergo2023buildingnormalizingflowsstochastic,albergo2023stochasticinterpolantsunifyingframework,lipman2022flow}. 
We then construct interpolating samples by
\begin{align}
    \label{eq:stoch_int}
    x_t = I(x_0,x_1,t),
\end{align}
where $x_0 \sim \rho_0,x_1 \sim \rho_1$, and $I$ satisfies $I(x_0,x_1,0)=x_0,I(x_0,x_1,1)=x_1$, e.g., $I(x_0,x_1,t) = (1-t)x_0+tx_1$ on $\mathbb R^n$ and the geodesic interpolation on $\mathbb T^{3}$ and $ SO(3)$. To learn the corresponding equilibrium potentials we employ score matching (SM).

\paragraph*{Denoising Score matching}
Denoising score matching (DSM) allows us to learn the score of a noised version of $\rho_t$
\begin{equation}
    \tilde \rho_t = \rho_t \star \mathcal N_{\sigma_t^2}.
\end{equation}
Performing TI along $\tilde \rho_t$  approximates TI along $ \rho_t$ as long as $\sigma_0\approx 0$ and $\sigma_1 \approx 0$. If $\sigma_0=\sigma_1=0$, then the boundary conditions are exactly satisfied, but DSM requires a non-zero $\sigma$, thus we choose $0<\sigma_0,\sigma_1 \ll 1$ only slightly violating the boundary conditions.
Conditioned on $x_0$ and $x_1$, $\tilde \rho_t(x_t|x_0,x_1)$ is an isotropic Gaussian with mean $I(x_0,x_t,t)$, variance $\sigma_t$ and Stein-score $\nabla \log \tilde \rho_t(x_t|x_0,x_1)=\frac{1}{\sigma_t}(I(x_0,x_1,t)-x_t)=-z/\sigma_t$. The marginal score $\nabla \log \tilde \rho_t(x)$ reads
\begin{equation}
    \nabla \log \tilde \rho_t(x) = \mathbb{E}_{x_0,x_1}\left[\nabla \log \tilde \rho_t(x_t|x_0,x_1)\right] = \mathbb{E}_{x_0,x_1}\left[-z/\sigma_t\right],
\end{equation}
enabling the training of an energy-based model using the usual DSM objective \citep{vincent2011connection}. This approach can be naturally extended to non-Euclidean $\mathcal M$ by learning the score in local charts on the manifold \citep{de2022riemannian}. 
\paragraph*{Target Score matching}
\citet{de2024target} recently proposed target score matching (TSM), an alternative SM objective that is well-suited for our application. TSM relies on having access to target score functions at the endpoints, $\nabla \log \rho_0 = -\nabla U_0 $ and $\nabla \log \rho_1 = -\nabla U_1 $. Decomposing  $\nabla \log \rho_t$, as
\begin{align}
    \nabla \log \rho_t(x_t)&=(1-t)^{-1}\int \nabla \log \rho_0(x_0) \rho(x_0,x_1|x_t) \mathrm dx_0 \mathrm dx_1 \\
    &= t^{-1}\int \nabla \log \rho_1(x_1) \rho(x_0,x_1|x_t) \mathrm dx_0 \mathrm dx_1, 
\end{align}
combined with the availability of the target scores, yields the objectives
\begin{align}
    \label{eq:TSM}
    \mathcal L^0_{\mathrm {TSM}} &= \mathbb E_{x_0,x_1,x_t} ||\nabla_x U_t(x_t)+\tfrac{1}{1-t}\nabla \log \rho_0(x_0)||^2 \\
    \mathcal L^1_{\mathrm {TSM}} &= \mathbb E_{x_0,x_1,x_t} ||\nabla_x U_t(x_t)+\tfrac{1}{t}\nabla \log \rho_1(x_1)||^2.
\end{align}
In our experiments we combine these two terms and work with a target score matching objective that always uses the closer endpoint of the $[0,1]$ interval,
\begin{equation}
    \mathcal L_{\mathrm {TSM}} = \mathbb I_{t<0.5}\mathcal L^0_{\mathrm {TSM}} +  \mathbb I_{t\geq 0.5}\mathcal L^1_{\mathrm {TSM}}.
\end{equation}

\paragraph*{Rolling estimate of $F_{0\to 1}$} Both SM objectives require the computation of the spatial gradient $\nabla_x U_t (x)$ at $(t,x)\sim(\mathcal U([0,1]),\rho_t)$. The temporal derivative of the energy function, $\partial_t U_t (x)$, can be easily calculated thanks to automatic differentiation at the same points, yielding a Monte Carlo estimate of the TI integral in Equation (\ref{eq_TI}),
\begin{equation}
    \label{eq:rolling_estimate}
    \Delta \hat F_{0\rightarrow 1} = \mathbb E_{t\sim\mathcal U([0,1]),x\sim \rho_t}\big[  \partial_t U_t(x)\big].
\end{equation}
In our experiments, at every training step we compute both the spatial and temporal derivatives of $U_t$  and use $\nabla U_t$ for the SM objective and $\partial_t U_t$ for a rolling estimate of the free-energy difference.

\subsection*{Solvation free energies}
We evaluate the proposed method by estimating solvation free energies, i.e., we concentrate on the case where the potentials $U_0, U_1$ are of the form
\begin{align}
    U_0 &= U_{\mathrm{solvent}} + U_{\mathrm{solute}} \\
    U_1 &= U_{\mathrm{solvent}} + U_{\mathrm{solute}}+ U_{\mathrm{solute-solvent}}, 
\end{align}
where $U_{\mathrm{solvent}}, U_{\mathrm{solute}}$ and $U_{\mathrm{solute-solvent}}$ represent the interactions within the solvent, within the solute and the cross interactions between the two components, respectively. In Section \ref{sec:experiments} we consider two experimental setups, ($i$) an LJ solute included in an LJ solvent and ($ii$) the inclusion of a rigid solute molecule in a solvent of rigid water molecules. We impose periodic boundary conditions, leading to configuration spaces $\mathbb T^{3}$ and $[\mathbb T^{3}\times SO(3)]^N$, respectively. The solute is rigid in both cases, implying $U_{\mathrm{solute}} \equiv 0$. As described below, most properties of these systems are shared. When discussing the differences between them we refer to the two setups by {LJ experiment} and {hydration experiment}.

\paragraph*{Interactions} We consider system where the components interact  via  Lennard-Jones (LJ) and Coulomb interactions. A softening of these interactions is necessary as the interpolating samples might bring particles close to each other, causing numerical issues with the unsoftened interaction. We use the LJ interaction with softening parameter\citep{BEUTLER1994529} given by 
\begin{align}
    \label{eq:soft_LJ}
    U(r,a)=4\varepsilon \left[\left(\frac{\sigma^2}{a\sigma^2+ r^2}\right)^{6}-\left(\frac{\sigma^2}{a\sigma^2+r^2}\right)^{3}\right],
\end{align}
where $r_{}$ is the distance between the particles and $(\epsilon,\sigma)$ denote the energy and length scales of the LJ interactions. When $a=0$, no softening is applied, and larger values of $a$ mean larger softening deformation of the original Lennard-Jones interactions. Note that equation (\ref{eq:soft_LJ}) corresponds to evaluating the unsoftened Lennard-Jones potential at $\sqrt{r^2 + a\sigma^2}$.  The parameters of the LJ interactions between different species, $A$ and $B$, of particles are computed by the Lorentz-Berthelot rules \citep{LorentzH.A.1881UdAd,berthelot1898melange},  $\varepsilon_{AB}=\sqrt{\varepsilon_A \varepsilon_B}, \sigma_{AB}=\frac{1}{2}(\sigma_A+\sigma_B)$. 

Coulomb interactions are modelled with the reaction field method \citep{tironi1995generalized}
\begin{align}
    U_{\text{Coulomb}}(r)&=
    \begin{cases} k_C{q_1q_2}\left(\frac{1}{r}+k_{\text{rf}}r^2-c_{\text{rf}}\right), \quad \text {if}  \quad r\leq r_{\text{cutoff}} \\
    0, \quad \text {if}  \quad r>r_{\text{cutoff}}
    \end{cases} \\
    k_{\text{rf}}&= \left(\frac{1}{r_\text{cutoff}^3}\right)\left(\frac{\epsilon_\text{solvent}-1}{2\epsilon_\text{solvent}+1}\right) \\
    c_{\text{rf}}&= \left(\frac{1}{r_\text{cutoff}}\right)\left(\frac{3\epsilon_\text{solvent}}{2\epsilon_\text{solvent}+1}\right),
\end{align}
where $r_\text{cutoff}$ is the cutoff radius of the Coulomb interaction, $k_C$ is Coulomb's constant, $q_1,q_2$ are the charges of the atoms, $r$ is the distance between them and $\epsilon_\text{solvent}$ 
is the dielectric constant of the solvent. To soften this potential with softening parameter $a$, we define
\begin{equation}
    \label{eq:soft_coulomb}
     U^\text{soft}_{\text{Coulomb}}(r,a)=  U_{\text{Coulomb}}(\sqrt{r^2+a}).
\end{equation}

\paragraph*{Ansatz for $U_t$}
We parameterize the interpolating potential in a way that satisfies the boundary conditions at $t \in \{0,1\}$ \citep{mate2023learning}.
\begin{align}
    \label{eq:ansatz}
    U_t(x) = \, b_t^A U_{\mathrm{solvent}}(x,a_t^A) 
    + b_t^B U_{\mathrm{solute}}(x,a_t^B) 
    +b^{AB}_t U_{\mathrm{solvent-solute}}(x,a^{AB}_t)
    +b_t U_t^\theta(x),
\end{align}
where  $U_t^\theta(x)$ is a neural network with trainable parameters. We enforce the boundary conditions of the softening $t \rightarrow a_t^A,a^{AB}_t,a_t^B$  and  coupling functions $t \rightarrow b_t^A,b^{AB}_t,b_t^B$ with the parametrization
\begin{align}
    b_t^{A} = [1-t(1-t)]e^{\beta_t^{\mathrm{A}}}& \quad 
    b_t^{B} = [1-t(1-t)]e^{\beta_t^{\mathrm{B}}} \\
    b_t^{\mathrm{AB}} = te^{t\beta_t^{\mathrm{AB}}}&  \quad
    b_t = t(1-t)e^{\beta_t} \\
    a_t^{\mathrm{A}} = t(1-t)e^{\alpha_t^{\mathrm{A}}}& \quad 
    a_t^{\mathrm{B}} = t(1-t)e^{\alpha_t^{\mathrm{B}}} \\
    a_t^{\mathrm{AB}} &= (1-t)e^{t\alpha_t^{\mathrm{AB}}}, 
\end{align}
where the $t\mapsto \alpha_t^*,t\mapsto \beta_t^*$ functions are learned by a multilayer perceptron with two hidden layers, each containing 256 hidden units, and swish activations \cite{hendrycks2023gaussianerrorlinearunits}. 
These functional forms ensure that  at $t=0$ and $t=1$ the solute and solvent interactions are fully coupled and no softening is applied to them, while the solvent--solute interactions are fully decoupled at $t=0$ and coupled at $t=1$. We predict such a set of parameters for each interaction separately.

\paragraph*{Interpolation}
We use the term constituent to refer to the solvent LJ particles in the case of the LJ solvent and to the rigid water molecules in the case of the water solvent. Given samples $x_0,x_1 \sim (\rho_0,\rho_1)$, we construct the function $I(x_0,x_1,t)$ in Equation (\ref{eq:stoch_int}) by  translating both $x_0$ and $x_1$ to center the solute and pair the solvent constituents using on the optimal pairing  with respect to the squared toroidal distance between the snapshots $x_0,x_1$.  In the case of water solvent, we consider the distances between the oxygen atoms.  Note that we compute the OT-pairings between the solvent constituents from two given snapshots $x_0,x_1$ and not between datapoints in a batch as proposed by \cite{tong_conditional_2023}. We keep the solute frozen throughout the interpolation at its state in $x_1$. During the interpolation, the positions of the solvent constituents move along the toroidal geodesics given by the OT-pairing. In case of the water solvent, the orientations of the water molecules are also interpolated using the geodesic interpolation on $SO(3)$. To further minimize the rotational transport cost we exploit the inner $\mathbb Z_2$-symmetry of the rigid water molecule. Applying this flip transformation does not change the physical system itself, only its numerical representation,  thus we compute the geodesic distances corresponding to the physically equivalent orientations and choose the one with the minimal geodesic distance. Geometrically speaking, this means that the configuration space of a single rigid water molecule can be reduced from $\mathbb T^3 \times SO(3)$ to $\mathbb T^3 \times SO(3)/\mathbb Z_2$.

\paragraph*{Architecture}
For the LJ system, we parametrize $U_t^\theta$ in equation (\ref{eq:ansatz}) with the same graph neural network as in \cite{mate2024neural} with node features encoding the particle type (solute or solvent) of the nodes. In the hydration experiment, to push the method further, we  write $U_t^\theta$ as a sum of pairwise interactions between all pairs of atoms in the system.
First, we  label each atom into one of the following four categories: (1) central atom of solute; (2) hydrogen in solute; (3) hydrogen in solvent; (4) oxygen in solvent. Then we consider the edge between all pairs of atoms, excluding pairs within the same molecule and label these edges into seven partitions, $\{P_1,...,P_7$\}, by the labels of their endpoints. We then predict an energy for each pair by using seven different multi-layer perceptrons, one for each category, taking as input the distance between the edge endpoints, the distance from the solute, and time as input,
\begin{equation}
    U_t^\theta = \sum_{k}\sum_{(i,j)\in P_k} f_{\psi_k}(t,d_{ij},d_i,d_j),
\end{equation}
where $k$ runs over the seven partitions of the edges, $t$ is time, $d_{ij}$ denotes the distance between atoms $i$ and $j$ and $d_i$ the distance between atom $i$ and the central atom of the solute. The functions $f_{\psi_k}$ with parameters $\psi_k$ are parametrized by multilayer perceptrons of with hidden layers $[96,96]$.

\paragraph*{Regularization of $\partial_t U_t$} The function we are aiming to learn is $U_t(x)$. The SM objective controls its behavior with respect to variations of $x$, but there is no training signal controlling the behavior of $U_t(x)$ with respect to changes to the temporal argument. To avoid large absolute values of $\partial_t U_t(x)$ we also include the $L_2$-norm of $\partial_t U_t$ to our optimization objective. Our training objective is thus
\begin{align}
    \label{eq:full_objective}
    \mathcal L(\theta) &= \mathbb E_{t\sim\mathcal U([0,1]),x_t\sim \rho_t}\Big[\lambda_t\mathcal L_{\mathrm{SM}}+ \kappa (\partial_t U_t(x_t))^2 \Big],
\end{align}
where $U_t$ depends on $\theta$ through equation \ref{eq:ansatz}, $ \mathcal L_{\mathrm{SM}} \in \{\mathcal L_{\mathrm{DSM}},\mathcal L_{\mathrm{TSM}}\}$, and $\lambda_t$ is the time-dependent weighting of the score matching term.
Since $\partial_t U_t$ is the very thing we are interested in measuring, we use a small relative weight for its regularization in our experiments, $\kappa \in \{10^{-8},10^{-10}\}$. The goal is not to minimize the $L_2$-norm of $\partial_t U_t$, but rather to prevent it from diverging.

\section{Results}
\label{sec:experiments}

\paragraph*{LJ solute in an LJ fluid}
\label{sec:experiments_LJ}

We consider a system of 512 particles interacting via an LJ potential. The LJ parameters of the solvent are denoted by $(\varepsilon_A, \sigma_A)$ while for the solute $(\varepsilon_B,\sigma_B)=(2\varepsilon_A,2\sigma_A)$. The simulation box is of size $9\sigma_A \times9\sigma_A \times 9\sigma_A$ with periodic boundary conditions. To train the model we employ denoising score matching.

\begin{figure*}[htb]
     \centering
     \begin{subfigure}
         \centering
         \includegraphics[width=0.63\textwidth]{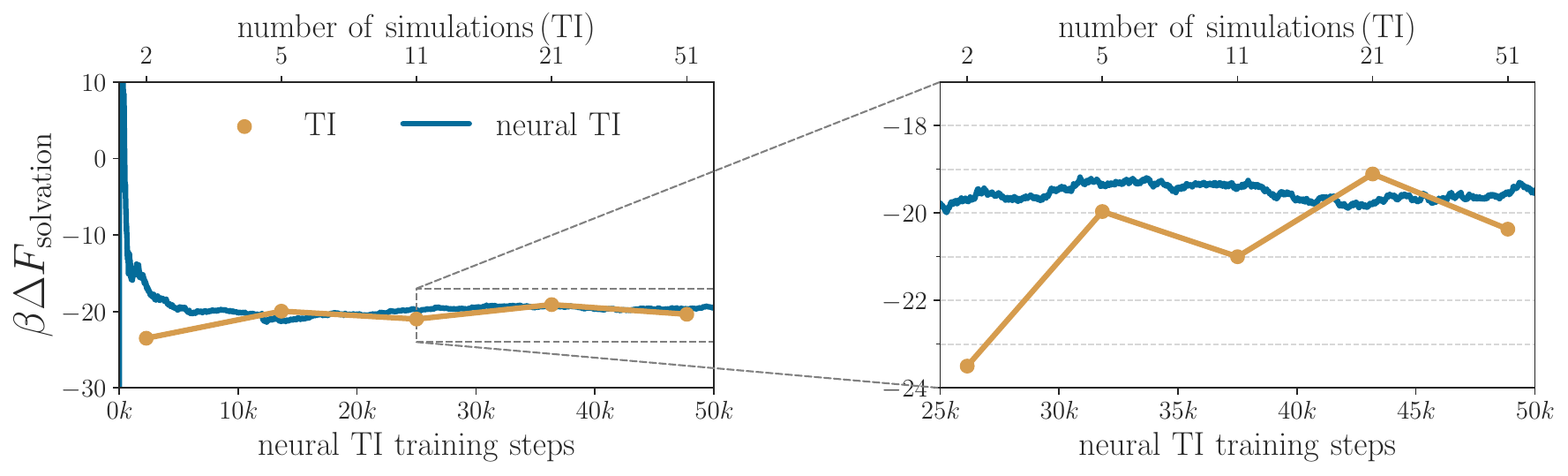}
     \end{subfigure}
     \hfill
     \begin{subfigure}
         \centering
         \includegraphics[width=0.34\textwidth]{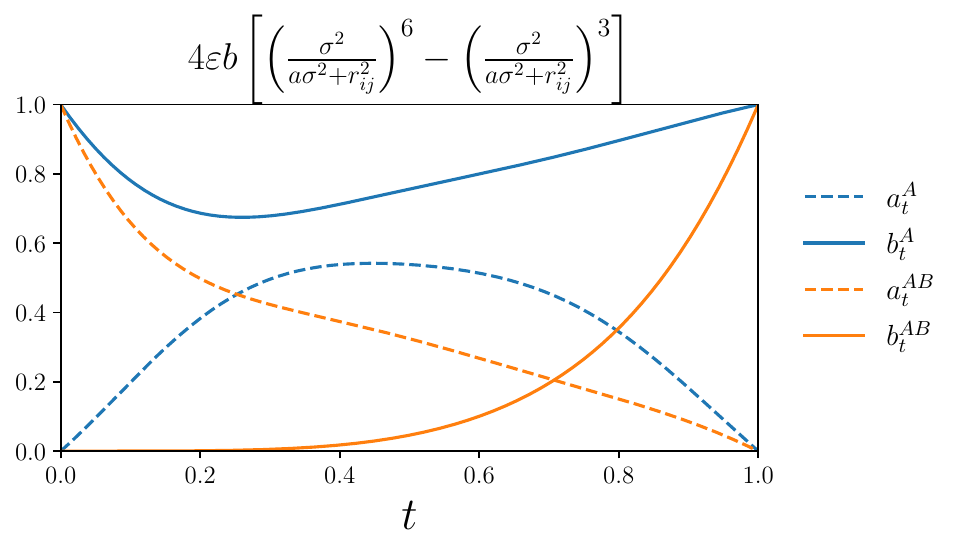}
     \end{subfigure}
     \caption{Solute solvation of a Lennard-Jones fluid. Left: Solvation free-energy estimates by TI and neural TI as a function of number of reference simulations and training steps, respectively. Center: Detailed view at later training stages. Softening parameters of the LJ interactions after training (right).}
     \label{fig:resultsLJ}
\end{figure*}

We compare our estimates with TI using evenly spaced time-slices along the time-interval $[0,1]$. We perform 51 simulations in total at $t \in \{0.00, 0.02, 0.04, ..., 1.00\}$, and find that the solvation free energy lies between $-21k_BT$ and $-19k_BT$. The rolling estimate of neural TI converges to approximately $-19.5 k_BT$. The time evolution of these values are shown in the left subplot of Figure \ref{fig:resultsLJ}. 
The solvation free energy estimate uses data only from the two endpoints and shows significantly lower variance than the associated standard TI computation, that requires between 5 and 50 interpolating timesteps. The right subplot of Figure \ref{fig:resultsLJ} displays the learned softening schedule of the solvent--solvent and solute--solvent Lennard-Jones interactions. The learned softening functions show that our interpolation does not correspond to the usual linear interpolation $(1-t) U_0 + tU_1$, even the interactions within the solvent are softened and decoupled along the trajectory. Consistent with our earlier work on neural TI \cite{mate2024neural}, we find that neural TI and traditional TI converge in a comparable amount of time.

\paragraph*{Hydration free energies}
\label{sec:experiments_hydr}
To move to more realistic chemical systems, we now extend the experimental setup to compute hydration free energies of small molecules, namely water and methane, from atomistic simulations. We work with rigid molecules, i.e., we freeze the bonded interactions. The configuration of a single molecule is thus described by the position of its central atom (i.e., oxygen for water, carbon for methane) and its orientation, i.e., a group element of $SO(3)$. The solvent consists of 216 TIP4P water molecules \citep{jorgensen1983comparison} at density 1 g/cm$^3$, resulting in a simulation box of size $18.62 \mathrm{\r{A}} \times 18.62 \mathrm{\r{A}} \times 18.62 \mathrm{\r{A}}$. The hydration of water---solvation of water in water---simply consists of coupling one extra TIP4P molecule in the box. For methane, we rely on the atomistic OPLS (OPLS-AA) force field \citep{mackerell1998all}. 
For the Coulomb interactions we use a cutoff radius  of $8.5 \mathrm{\r{A}}$ and dielectric constant $\epsilon_\text{water}=78$. Importantly, unlike the usual practice of TI, we do not prescribe to couple the Lennard-Jones interactions first, and the Coulombic interactions later---we instead let the network learn the coupling schedule with separate softening parameters for the two interactions. To train the model we employ target score matching.

\begin{figure*}[htb]
     \centering
     \begin{subfigure}
         \centering
         \begin{tikzpicture}
        \node[draw=black, dashed, thick, inner sep=0pt] (img) {\includegraphics[width=0.25\textwidth,trim={2cm 2.5cm 2cm 2.5cm},clip]{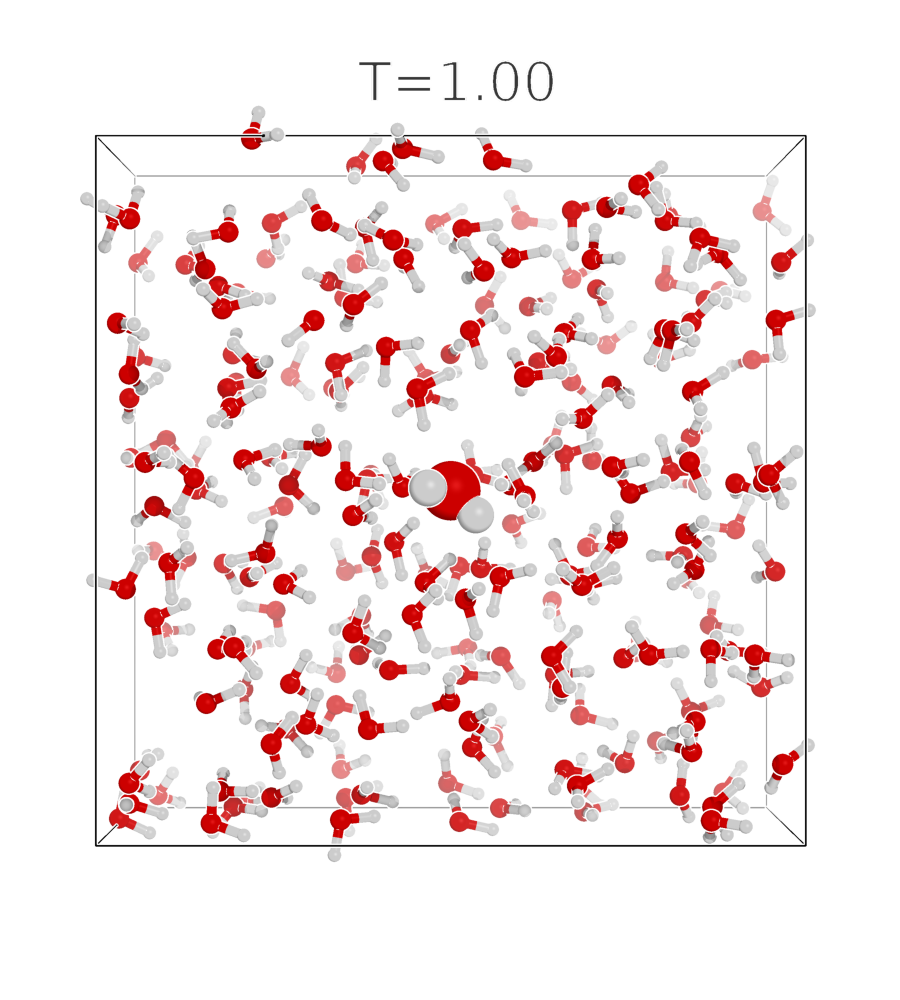}};
        \node[below=0.1cm of img] {};
        \end{tikzpicture}
     \end{subfigure}
     \hfill
          \begin{subfigure}
         \centering
         \begin{tikzpicture}
        \node[draw=black, dashed, thick, inner sep=0pt] (img) {\includegraphics[width=0.25\textwidth,trim={2cm 2.5cm 2cm 2.5cm},clip]{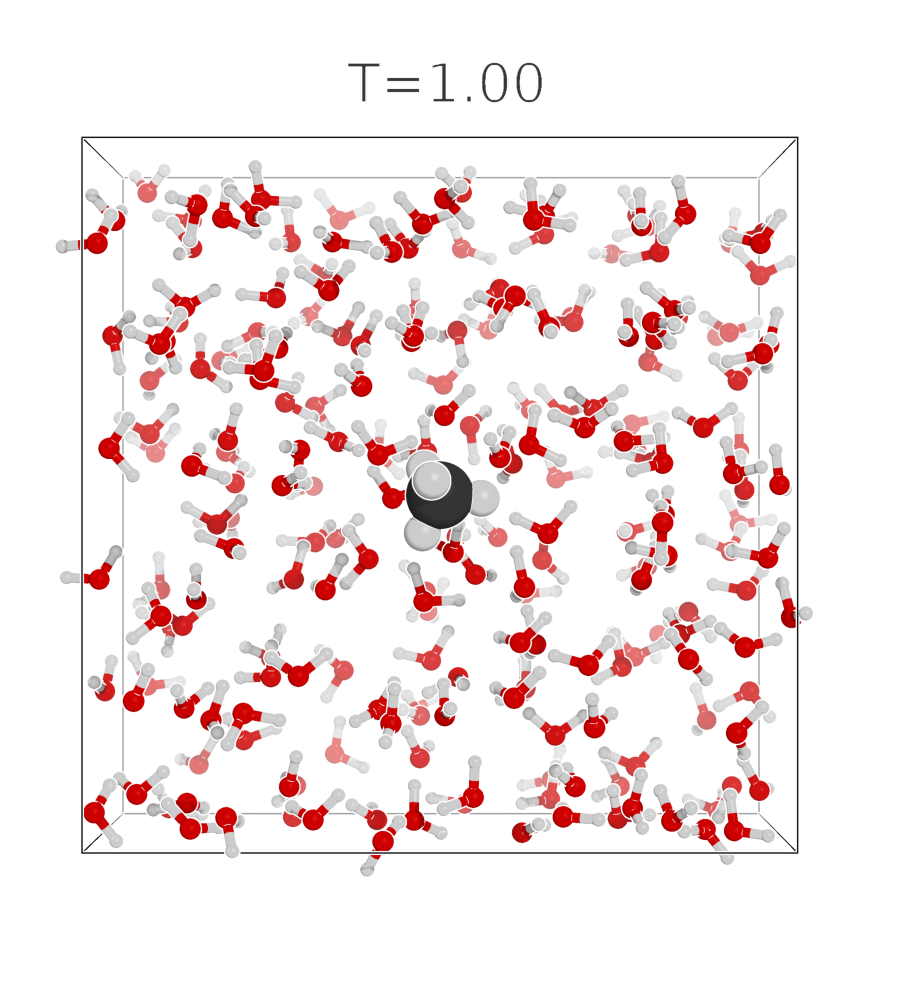}};
        \node[below=0.1cm of img] {};
        \end{tikzpicture}
     \end{subfigure}
     \hfill
     \begin{subfigure}
         \centering
\includegraphics[width=0.43\textwidth,trim={0cm .25cm .2cm .2cm},clip]{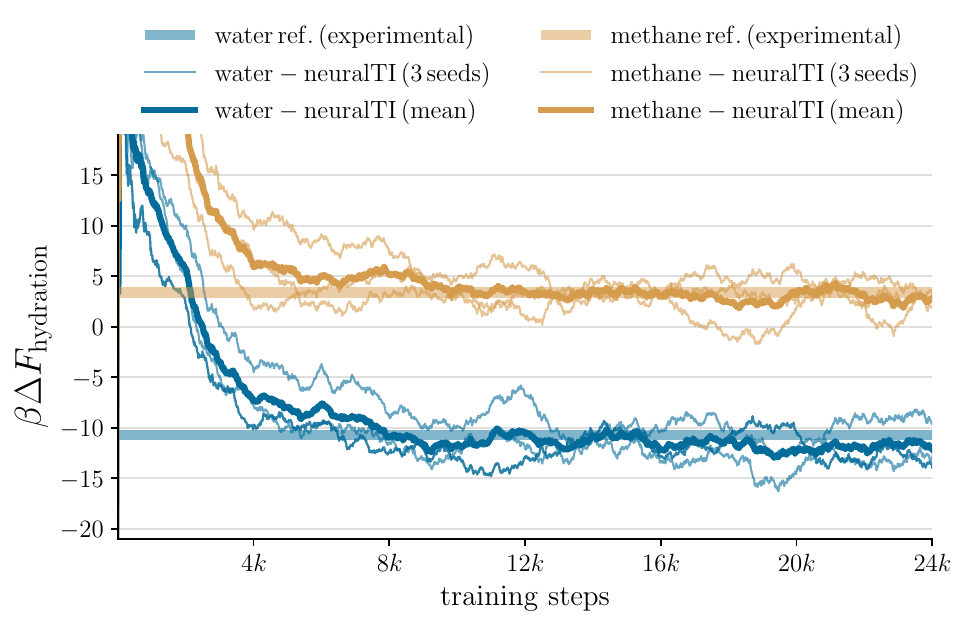}
     \end{subfigure}
     \caption{Illustration of the coupled water (left) and methane (center) molecules. Hydration free-energy estimates of water and methane in TIP4P water (right). The dashed lines denote the experimental hydration free energies and the the curves of the same our rolling estimate from three different random seeds for each solute.}
    \label{fig:results_hydration}
\end{figure*}

We benchmark our calculations on previously reported calculations of the hydration free energy for both water and methane: $-10.7 k_BT$ \citep{jorgensen1989free} and $3.4k_BT$ \citep{mobley2009small}, respectively. Figure \ref{fig:results_hydration} (right panel) compares our rolling neural TI estimates (equation (\ref{eq:rolling_estimate})) with the reference values mentioned. For each system we provide 3 individual calculations with different initial random seeds. For the moving average we use a window size of 2,000 batches of batch size 24, i.e., 48,000 samples. The runs for water and methane use the same architecture of the neural network potential, same configuration, the runs only differ by the solute. The runs for the water and methane visibly diverge after 2,000 training steps and the means over the 3 respective seeds stay in the $\pm 2 k_BT$ range of the experimental value after 10,000 training steps. The standard error of the mean for both systems is below $3k_BT$.  All runs were performed on an Nvidia 3090 and took approximately three hours to reach the reported 24,000 training steps.

Given that the two solutes are of comparable size and complexity, we may assume that two systems, and in particular their solute--solvent coupling process, share some similarities. In the ideal case, a model trained on one of the solutes could give a reasonable prediction when evaluated on the other. We test this with the trained models and confirm that this is indeed the case: both models learn features that are useful when modeling the coupling of the other solute and predict the hydration free energy of the unseen solute within a $4k_BT$ accuracy. 
We attribute this transferability to our ansatz (equation \ref{eq:ansatz}), which enforces the target Hamiltonians, along with the contribution of solvent-solvent interactions. The learned softening of the endpoints and the accurate modeling of solvent-solvent interactions throughout the coupling process are likely key factors underlying the observed transferability.
All numerical values are reported in Table \ref{tab:results}.
\begin{table}[h]
\centering
\renewcommand{\arraystretch}{1.5}
\caption{Experimental and neuralTI-estimated hydration free energies. All the neuralTI estimates are computed over $64,000$ $(t,x_t)$ samples from $(\mathcal U([0,1]),\rho_t)$ and averaged over three seeds. The colored cells contain the transfer predictions, i.e., predictions on solutes that the model was not trained on.}
\label{tab:results}
\begin{tabular}{ccc}
\toprule
 \textbf{} & $\Delta F_{\text{hydration}}^{\text{water}}$  & $\Delta F_{\text{hydration}}^{\text{methane}}$ \\ \midrule
experimental reference value & $-10.7 k_BT$  & $3.4 k_BT$  \\ 
neural TI with a model trained on water & $-11.5 k_BT$  &  \cellcolor{red!20}  $1.6 k_BT$  \\
neural TI with a model trained on methane &  \cellcolor{red!20} $-13.8 k_BT\,\,\,\,$  & $3.7 k_BT$  \\
\bottomrule
\end{tabular}

\end{table}

\section{Conclusion}
We build on ideas introduced by \citet{mate2024neural} to estimate free-energy differences between arbitrary pairs of target potentials by performing thermodynamic integration with a neural-network potential. Motivated by generalizations of diffusion models \cite{lipman2022flow,albergo2023buildingnormalizingflowsstochastic}, our method relies on two, possibly non-trivial, Hamiltonians over the same configuration space, samples from their Boltzmann distributions, and an interpolating procedure between samples from these two distributions. The upside of this approach, compared to traditional TI, is that we only need reference Boltzmann distributions for the end points, entirely alleviating the need for intermediate reference MC or MD simulations. Additionally, we simplify and improve the neuralTI approach by incorporating techniques like the rolling estimate of free-energy differences (equation \ref{eq:rolling_estimate}), target score matching, and the regularization of the temporal derivative (equation \ref{eq:full_objective}), each of which we found beneficial during our experiments.

The method is verified by accurately estimating the solvation free energy of a Lennard-Jones solute in a Lennard-Jones solvent using only data from the endpoint simulations, matching the accuracy of standard TI, which typically requires tens of intermediate simulations. We also predict the hydration free energies of rigid water and methane, showing good agreement with experimental values, using a simple three-body potential. Importantly, we do not observe or claim a computational speedup compared to traditional TI. We do however see promise for a computational advantage by learning an interpolating potential that is transferable between different systems. We observe a simple instance of such a transferable potential between water and methane in our hydration experiment (Table \ref{tab:results}).

Future work could scale the method to more complex systems. The main limitation of the current approach is the rigidity assumption of the molecules. Overcoming this requires introducing bonded interactions, which removes the need for the 
$SO(3)$ factors in the configuration space but also introduces forces acting on different length scales, potentially posing numerical challenges.  Additionally, it would be of interest to control the variance of the TI estimate (eq. \ref{eq:rolling_estimate}) without explicitly regularizing it (eq. \ref{eq:full_objective}).

\paragraph*{Code availability}
Our implementation is available at the following repository:
\url{https://github.com/balintmate/neuralTI-solvation}.


\section*{Acknowledgments}
We thank Aleksander Durumeric, Daniel Nagel, Lorenz Richter, and Youssef Saied for discussions. BM acknowledges financial support by the Swiss National Science Foundation under grant number CR - SII5 - 193716 (RODEM). TB acknowledges support by the Deutsche Forschungsgemeinschaft (DFG, German Research Foundation) under Germany's Excellence Strategy EXC 2181/1 - 390900948 (the Heidelberg STRUCTURES Excellence Cluster).

\bibliography{main}

\end{document}